\documentclass[preprint]{aastex}
\usepackage{emulateapj5}
\usepackage{apjfonts}
\usepackage{natbib}

\input psfig.tex

\def\aa{{A\&A}}

\def\aj{{AJ}}

\def\amin{$^\prime$}
\def\annrev{{ARA\&A}}
\def\apj{{ApJ}}
\def\apjs{{ApJS}}
\def\asec{$^{\prime\prime}$}

\def\deg{$^{\circ}$}

\def\e#1{$\times$10$^{#1}$}

\def\flux{erg s$^{-1}$ cm$^{-2}$}

\def\fsec{\hbox{$.\mkern-4mu^s$}}

\def\kms{km s$^{-1}$}

\def\lax{{$\mathrel{\hbox{\rlap{\hbox{\lower4pt\hbox{$\sim$}}}\hbox{$<$}}}$}}
\def\gax{{$\mathrel{\hbox{\rlap{\hbox{\lower4pt\hbox{$\sim$}}}\hbox{$>$}}}$}}
\def\simlt{\lower.5ex\hbox{$\; \buildrel < \over \sim \;$}}
\def\simgt{\lower.5ex\hbox{$\; \buildrel > \over \sim \;$}}
\def\lum{erg s$^{-1}$}

\def\mnras{{MNRAS}}
\def\nat{{Nature}}
\def\pasp{{PASP}}

\def\percm2{cm$^{-2}$}
\def\percm3{cm$^{-3}$}

\def\solmass{$M_\odot$}

\slugcomment{To appear in {\it The Astrophysical Journal (Letters)}.}
\shorttitle{ACCRETION LUMINOSITY OF BLACK HOLE IN M15}
\shortauthors{HO ET AL.}

\begin{document}

\title{A Stringent Limit on the Accretion Luminosity of the Possible 
Central Black Hole in the Globular Cluster M15}

\author{Luis C. Ho\altaffilmark{1}, Yuichi Terashima\altaffilmark{2,3}, 
and Takashi Okajima\altaffilmark{4}}

\altaffiltext{1}{The Observatories of the Carnegie Institution of Washington, 
813 Santa Barbara St., Pasadena, CA 91101-1292.}

\altaffiltext{2}{Department of Astronomy, University of Maryland, College 
Park, MD 20742-2421.}

\altaffiltext{3}{Institute of Space and Astronautical Science, 3-1-1 Yoshinodai, Sagamihara, Kanagawa 229-8510, Japan.}

\altaffiltext{4}{NASA Goddard Space Flight Center, Code 662, Greenbelt, MD 
20771.}

\begin{abstract}
The globular cluster M15 has recently been found to host a possible central 
black hole with a mass of $\sim$2000 \solmass.  A deep, high-resolution 
{\it Chandra}\ image failed to detect the ``nucleus'' of the cluster in 
X-rays.  The upper limit on the X-ray luminosity ($L_{\rm x}$ \lax\ 
$5.6 \times10^{32}$ \lum) corresponds to a bolometric Eddington ratio of 
$L_{\rm bol}/L_{\rm Edd}$ \lax\  $(2-4) \times 10^{-8}$.  Combining this limit 
with an estimate of the electron density of the intracluster ionized plasma 
derived from pulsar dispersion measures, we show that the radiative efficiency 
of the accretion flow, if it accretes at the Bondi rate, must be much lower 
than that of a standard optically thick, geometrically thin disk.  
\end{abstract}


\section{Introduction}

The discovery of X-ray sources in Galactic globular clusters nearly 30 years 
ago (Giacconi et al. 1974; Canizares \& Neighbours 1975; Clark, Markert, \& Li 
1975) prompted speculation that central massive ($\sim 100-1000$ \solmass) 
black holes (BHs) may exist in these systems (Bahcall \& Ostriker 1975; Silk \& 
Arons 1975).  In this respect, M15, one of the densest globular clusters 
detected in X-rays, historically has been scrutinized most thoroughly (see 
review by van~der~Marel 2001).  Often regarded as a prototypical 
core-collapsed globular cluster (Djorgovski \& King 1986), M15 has a projected 
stellar density profile that rises steeply toward the center (Guhathakurta 
et al. 1996; Sosin \& King 1997), seemingly in excellent agreement with models 
for the distribution of stars around a massive collapsed object (Peebles 1972; 
Bahcall \& Wolf 1976).  Unfortunately, core collapse of the cluster, induced 
by two-body relaxation, can produce density profiles that closely mimic 
those generated by the influence of a BH (Grabhorn et al. 1992), and thus the 
photometric evidence for a BH remained ambiguous.  Kinematic data can provide a 
much more definitive probe of the central mass distribution.  However, 
despite numerous efforts (e.g., Peterson, Seitzer, \& Cudworth 1989; Dubath \& 
Meylan 1994; Gebhardt et al. 1994, 1997; Dull et al. 1997; Drukier et al. 
1998), the ground-based spectroscopic studies of M15 have met with only 
limited success.  This is largely due to the tremendous difficulty of 
measuring radial velocities for individual stars in the crowded cluster core, 
even under conditions of exceptional seeing.  As summarized in the latest 
study by Gebhardt et al. (2000), the central kinematics of M15 are consistent 
with the presence of a 2500 \solmass\ dark object, but the data can be equally 
well fitted with a mild, not unreasonable, increase in the mass-to-light ratio 
toward the center.  

A major breakthrough was recently achieved in the search for the elusive 
central BH in M15.  Using a series of longslit spectra taken with the {\it 
Hubble Space Telescope}, van~der~Marel et al. (2002) and Gerssen et al. (2002, 
2003) significantly increased the sample of radial velocities measured within 
the cluster core, thereby permitting a more robust analysis of the central 
kinematics. The data indicate the presence of a central dark mass, although its 
nature is still unclear.  Models with a BH mass of $\sim$2000 \solmass\ provide 
a marginally better fit to the data than those without one, but not at a 
statistically significant level of confidence (Gerssen et al. 2002, 2003).  The 
dynamical simulations of Baumgardt et al. (2003) show that a central 
concentration of non-luminous, massive stellar remnants can account for the 
data equally well.  This explanation, however, is uncertain because it 
depends on the assumption that all the neutron stars are retained in the 
cluster  (Baumgardt et al. 2003; Gerssen et al. 2002, 2003).  Thus, a massive 
BH cannot be ruled out in M15.  In a parallel study, Gebhardt, Rich, \& Ho 
(2002) announced the detection of a 2\e{4} \solmass\ BH in G1, a luminous
globular cluster in the galaxy M31.  This finding lends additional confidence 
to the BH interpretation for the case of M15.

The possible existence of central massive BHs in star clusters opens up many 
new avenues of investigation.  Particularly interesting is the prospect of 
probing accretion physics in the ``active nucleus'' of the cluster, by direct 
analogy with the study of active galactic nuclei in external galaxies.  
Nearby stellar clusters potentially offer a completely fresh vantage point for 
studying nuclear activity.  In contrast to extragalactic nuclei, Galactic 
clusters are well resolved into individual stars, and their structure and 
dynamics are much better understood.

This paper reports sensitive, high-resolution {\it Chandra}\ observations of 
the central region of M15.  The nucleus is not detected in X-rays, down to an 
exceedingly stringent limit of $\sim2.2\times 10^{-9}$ of the Eddington 
luminosity of a 2000 \solmass\ BH.  We combine this measurement with estimates 
of the central gas density to constrain the radiative efficiency of the 
accretion flow.

\section{Observations and Results}

Our analysis is based on archival data (PI: J.~E. Grindlay) acquired with 
{\it Chandra}\ (Weisskopf, O'Dell, \& van~Speybroeck 1996) using the High 
Resolution Camera (HRC-I; Murray et al. 1997).  HRC-I has a field of view of 
30\amin$\times$30\amin\ and a pixel scale of 0\farcs13, which well samples the 
point-spread function (PSF) of the telescope (FWHM $\approx$ 0\farcs4).  
There are a total of three observations, performed on 2001 July 13, August 3, 
and 

\vskip 0.3cm

\psfig{file=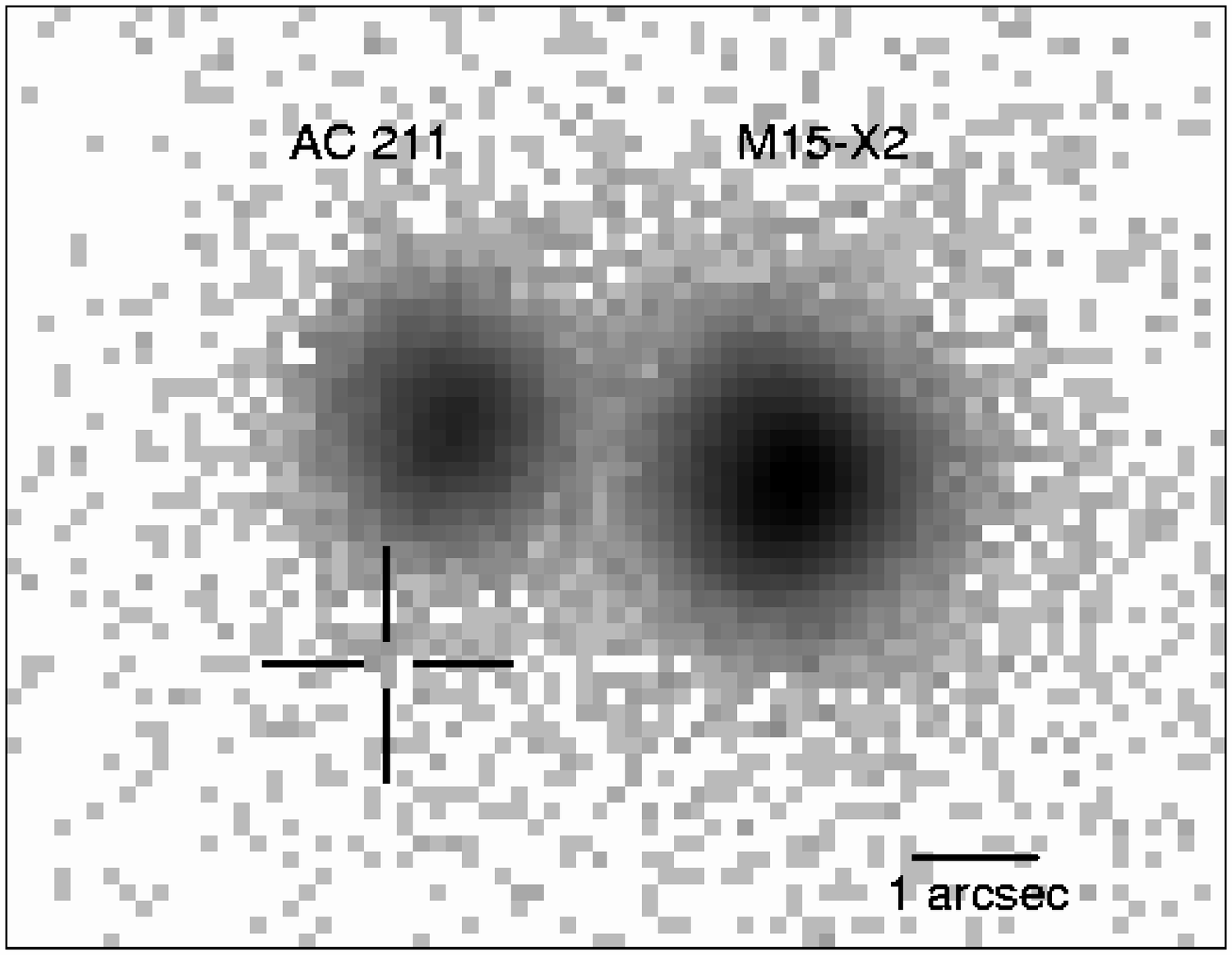,width=8.5cm,angle=0}
\figcaption[fig1.ps]{
HRC-I image of a 7\farcs8$\times$10\asec\ region near the center of M15.
The two bright LMXBs are labeled, along with the optical position of the
nucleus of the cluster.  North is up and East is to the left.
\label{fig1}}
\vskip 0.3cm

\noindent
August 22.  The effective exposure times are 9.1, 8.8, and 10.8 ks,
respectively.  We analyzed the data sets separately, and then combined.

Figure~1 shows the central 7\farcs8$\times$10\asec\ of the HRC-I image.  As 
recently demonstrated by White \& Angelini (2001) using a {\it Chandra}/ACIS 
image, the X-ray source 4U~2127+119 in fact consists of two sources 
separated by 2\farcs7: AC~211, a low-mass X-ray binary (LMXB) with a 
previously known bright optical counterpart, and M15 X-2, another LMXB that 
is coincident with a faint, blue star ($U \approx 19$ mag).  The {\it Chandra}\
positions of the two bright sources were compared with previous measurements 
[radio position of AC211 from Kulkarni et al. (1990); X-ray position of M15 
X-2 from White and Angelini (2001)] and found to be offset by $\sim$0\farcs2.
We shifted the images by this amount to align the X-ray and radio positions of 
AC211.  

The position of the cluster center (Gerssen et al. 2002), $\alpha\, =\, 
21^h 29^m 58\fsec335$, $\delta$ = 12\deg10\amin0\farcs89 (J2000), shows no 
significant counts in excess of the extended wings of the PSF from AC~211.  
This is illustrated quantitatively in Figure~2, which gives a one-dimensional 
projection of the surface brightness along the cluster center and AC211.  
The profile, derived by summing over a width of 6 pixels (0\farcs78), is 
used to model the shape of the PSF of AC211 and to calculate an upper limit 
on the nuclear flux. The choice of the width affects the following results only
slightly.  For both the first and second observations, the profile of AC211 is 
well fitted with a model consisting of a Gaussian and a Lorentzian, 
representing the core and the extended wings of the PSF, respectively.  We 
note that this model is not an accurate representation of the PSF for data 
with very good photon statistics. Indeed, in the third observation AC211 is 
about four times brighter than in the first and second observations, and 
systematic errors in the PSF model dominate over photon statistics.  In 
order to use our simple PSF model, we concentrate only on the first two 
observations.

We calculated an upper limit on the count rate from the nucleus by adding a 
Gaussian, whose width was fixed to the value determined from AC211, to model 
the emission from the cluster center.  Since the core of the PSF is well 
described by a Gaussian, a single Gaussian provides a reasonable description 
of the PSF shape for the very faint central source. We employed a maximum 
likelihood method in the fits because of the small number of counts in the 
tail of the profile. The errors on 

\vskip 0.3cm

\psfig{file=fig2.ps,width=8.5cm,angle=-90}
\figcaption[fig2.ps]{
A spatial cut centered on AC211.  The cluster center is located at pixel
$-15.7$.  The solid line is the best fit for the PSF of AC211, modeled as the
sum of a Gaussian and a Lorentzian.
\label{fig2}}
\vskip 0.3cm

\noindent
the counts were estimated using the 
approximation of Gehrels (1986).
 
Since M15 X-2 is roughly equidistant from the cluster center and the opposite 
side of AC211 (right-hand side in Fig.~2), the tail of its PSF should 
contribute roughly equally to both positions, provided that the PSF does not 
vary strongly with azimuth.  We examined the azimuthal angle dependence of the 
PSF by using an observation of 3C~273.  We find that any variation is at 
most $\pm$10\% (including statistical error) at 2\asec\ from the PSF peak, 
approximately the distance between AC211 and the cluster center.

We repeated these measurements for the first, second, and the combined
profile of the two observations.  The Gaussian component for the cluster
center was not required statistically in any of the fits. The 90\%
confidence upper limit (one parameter of interest) on the count rates
are 0.0059, 0.0043, and 0.0033 counts s$^{-1}$, respectively.  To convert the 
limit on the count rate to an X-ray flux, we assumed (1) that the spectrum 
between 0.2 and 10 keV can be described by a power law with a photon index of 
2.3, a spectral model that fits well the quiescent nucleus of the Galaxy 
(Baganoff et al. 2001) and M32 (Ho, Terashima, \& Ulvestad 2003), (2) 
that the line-of-sight absorbing column $N_{\rm H} = 5.8\times10^{20}$ 
cm$^{-2}$, calculated from $E(B-V) = 0.10$ mag (Harris 1996) and the relation 
$N_{\rm H}$ = 5.8\e{21} $E(B-V)$ cm$^{-2}$ (Savage \& Mathis 1979), and (3) 
the latest effective area for the combination of the high-resolution 
mirror assembly and the HRC (version 2.1).  These assumptions give $F_{\rm x} 
< 4.4 \times 10^{-14}$ \flux\ for the combined observations, or $L_{\rm x} 
< 5.6\times10^{32}$ \lum\ assuming a distance of 10.4 kpc (Harris 1996).

\section{Implications}

The nucleus of M15, if presumed to host a $\sim 2000$ \solmass\ BH, is 
extremely inactive.  The upper limit on the X-ray luminosity derived from the 
{\it Chandra}\ observations corresponds to an Eddington ratio of 
$L_{\rm x}/L_{\rm Edd}$ \lax\  2.2\e{-9}, where $L_{\rm Edd}\,=\,1.26 \times 
10^{38} \left(M/M_{\odot}\right)$ \lum.  Judging from the broad-band spectral 
energy distributions of galactic nuclei over a wide range of activity 
levels (Elvis et al. 1994; Ho 1999), the bolometric correction for the 
X-ray band is $\sim 7-20$.  Hence, $L_{\rm bol}/L_{\rm Edd}$ \lax\ 
$(2-4)\times 10^{-8}$.

Why is the nuclear BH of M15 so quiescent?  One possibility is that the 
cluster is completely devoid of gas.  Although post-main sequence stars 
continuously inject mass into the intracluster medium through mass loss, 
a variety of mechanisms can effectively remove it (e.g., Frank \& Gisler 1976; 
Spergel 1991; Smith 1999).  Indeed, nearly all efforts to detect gas, be it 
neutral or ionized, in globular clusters have resulted in null detections (see 
Knapp et al. 1996, and references therein).  While low, however, the gas 
content is not zero, and it would be of interest to estimate how much 
accretion luminosity one might expect to see.  Dispersion measures derived 
from radio observations of pulsars provide the most sensitive probe of the 
line-of-sight integrated column density of free electrons in globular clusters.
From an analysis of the population of millisecond pulsars in 47 Tucanae and 
M15, Freire et al. (2001) find an electron density of $n_e \approx 0.1$ and 0.2 
cm$^{-3}$, respectively.  We expect the plasma, which has been photoionized by 
the ultraviolet radiation field from post-asymptotic giant branch stars, to 
have a temperature $T_e \approx 10^4$ K.  To first order, this temperature is
consistent with the value derived assuming that the gas is in virial 
equilibrium with the stars.  For a central stellar velocity dispersion of 
$\sigma = 14$ \kms\ (Gerssen et al. 2002), $T = \mu m_p \sigma^2/k \approx 
3 \times 10^4$ K, where $\mu$ is the mean atomic weight, $m_p$ is the proton 
mass, and $k$ is Boltzmann's constant.

Let us assume that low-angular momentum gas in the vicinity of the BH accretes 
spherically, as described by Bondi (1952). The gravitational potential of the 
BH dominates the dynamics of the gas within the accretion radius, 
$R_{\rm a} \approx GM/c_{\rm s}^2$, where $c_{\rm s} \approx 0.1 T^{1/2}$ 
\kms\ is the sound speed of the gas.  For $T_e = 10^4$ K and $M$ = 2000 
\solmass, $c_{\rm s} \approx 10$ \kms\ and $R_{\rm a} \approx 0.1$ pc  
($\sim$2\asec\ at the distance of M15).  From the continuity equation, the 
Bondi accretion rate 
$\dot M_{\rm B}\,=\,4\pi R_{\rm a}^2 \rho_{\rm a} c_{\rm s}$, where
$\rho_{\rm a}$ is the gas density at $R_{\rm a}$.   Expressed in terms of
parameters appropriate for M15,

$$\dot M_{\rm B}\,\approx\,4.8\times 10^{-9}\,
\left(\frac{M}{2000\, M_{\odot}}\right)^2
\left(\frac{n}{0.2\, {\rm cm}^{-3}}\right)
\left(\frac{10\, {\rm km\,s}^{-1}}{c_{\rm s}}\right)^3
\,M_{\odot}\,{\rm yr}^{-1}.$$

\noindent
If this emission is produced by an optically thick, geometrically thin disk 
(Shakura \& Sunyaev 1973), the accretion luminosity is $L_{\rm acc} = \eta 
\dot M c^2 \approx 3\times10^{37} \left(\eta/0.1\right)$ \lum, where we have 
assumed a canonical radiative efficiency of $\eta = 0.1$ and $\dot M = 
\dot M_{\rm B}$.  This high value of the accretion luminosity clearly 
contradicts the observational upper limit of $L_{\rm bol} \approx (4-11) 
\times 10^{33}$ \lum\ (the range corresponds to an X-ray bolometric correction 
of 7--20), a factor of $\sim 3000-8000$ lower.  Unless the plasma density has 
been overestimated by 3--4 orders of magnitude, we are forced to conclude 
that either $\eta \ll 0.1$ or $\dot M \ll \dot M_{\rm B}$.  Both, in fact, may 
hold naturally in globular clusters, given their low gas content.  

Optically thin advection-dominated accretion flows (ADAFs; for a general 
overview, see Narayan, Mahadevan, \& Quataert 1998), which are radiatively 
inefficient, are thought to develop when accretion rates drop below a critical 
threshold of $\dot M_{\rm crit} \approx \alpha^2 \dot M_{\rm Edd} \approx 0.1 
\dot M_{\rm Edd}$, where the Eddington accretion rate is defined by 
$L_{\rm Edd}\,=\,\left(\eta/0.1\right) \dot M_{\rm Edd} c^2$ and the Shakura 
\& Sunyaev (1973) viscosity parameter is taken to be $\alpha \approx 0.3$ 
(Narayan et al. 1998).  From the above estimate of the Bondi accretion rate, 
$\dot M_{\rm B} \approx 10^{-4} \dot M_{\rm Edd}$, and so an ADAF, or 
some variant thereof (see Quataert 2001 for a review of the recent 
modifications of the basic ADAF model that incorporate the effects of outflows 
and convection), very likely exists in M15.  We believe this accounts for the 
extraordinary quiescence of its nucleus, should it truly host a central 
massive BH.  Similar arguments have been advanced to explain the dimness of 
supermassive ($\sim 10^6-10^9$ \solmass) BHs in the nuclei of nearby giant 
elliptical (e.g., Fabian \& Rees 1995; Mahadevan 1997) and spiral (Ho 2003) 
galaxies. 

While accretion flows in globular clusters are expected to be radiatively 
inefficient, how low can $\eta$ be?  Is $\eta$ \lax\ $10^{-4}$ possible?  
This issue is not yet well understood theoretically, as it depends on the 
currently uncertain physics of particle heating and acceleration (Quataert 
2001).  One can allow higher values of $\eta$ by reducing $\dot M$ below 
the Bondi rate.  Recent {\it Chandra}\ studies of elliptical galaxies 
find that their cores often accrete significantly below the Bondi rate 
(e.g., Loewenstein et al. 2001; Ho et al. 2003).  The accretion rate is 
thought to be curtailed by dynamical processes inherent to radiatively 
inefficient flows, which have the propensity to develop outflows 
(Blandford \& Begelman 1999) and convection (Narayan, Igumenshchev, \& 
Abramowicz 2000; Quataert \& Gruzinov 2000; Igumenshchev \& Narayan 2002).

Following Grindlay et al. (2001), we can turn the problem around and use 
the X-ray null detection to place an upper limit on the mass of the central 
BH.  Assuming again that the BH accretes at the full Bondi rate and that 
$\eta = 10^{-4}$, the upper limit on the bolometric luminosity translates 
into an upper limit of $\sim 600-1000$ \solmass\ for the central BH.

\section{Summary}

We use a sensitive, high-resolution {\it Chandra}\ image to place a stringent 
upper limit on the accretion luminosity of the nucleus of the globular cluster 
M15, which plausibly hosts a massive ($\sim$2000 \solmass) BH.  The bolometric 
luminosity of the nucleus is less than $(2-4)\times 10^{-8}$ of the Eddington 
luminosity of the BH.  If the central BH accretes via a standard optically 
thick, geometrically thin disk at the Bondi rate, which we calculate from the 
electron density of the intracluster ionized plasma derived from pulsar 
dispersion measures, the nucleus should be $\sim3-4$ orders of magnitude 
more luminous than observed.  Unless the accretion rate has been severely
overestimated, this fundamental inconsistency leads to the conclusion that the 
accretion process in M15 must be extremely radiatively inefficient, as 
theoretically predicted in the context of advection-dominated accretion flows 
or related models.  The constraint on the radiative efficiency is not yet 
precise, however, because of the uncertain influence of outflows or convection 
on the accretion rate.  Of course, the lack of an X-ray nucleus in M15 in 
itself can be taken as evidence that it does not contain a massive BH.  This 
interpretation remains viable until the nature of its central dark mass 
concentration can be established with greater certainty.

\acknowledgements
We acknowledge the contribution of J.~E. Grindlay, the PI of the original 
{\it Chandra}\ observations on which our analysis is based.  The research of 
L.~C.~H. is supported by the Carnegie Institution of Washington and by NASA 
grants from the Space Telescope Science Institute (operated by AURA, Inc., 
under NASA contract NAS5-26555).  T.~O. and Y.~T. are supported by the Japan 
Society for the Promotion of Science.  We thank an anonymous referee for 
providing constructive comments on the manuscript.

%

%

%
%
%


\end{document}